\documentclass[10pt]{article}
\pdfoutput=1
\usepackage{graphics, color}
\usepackage{graphicx}
\usepackage{amsmath}
\usepackage{amssymb}
\usepackage{amsfonts}
\usepackage{tensor}
\usepackage[usenames,dvipsnames]{xcolor}
\usepackage[normalem]{ulem}
\usepackage[bottom]{footmisc}

\definecolor{orange}{rgb}{1,0.5,0}
\definecolor{grey}{rgb}{.5,.5,.5}
\definecolor{bluegreen}{rgb}{0,.5,.5}
\definecolor{darkgreen}{rgb}{0,.5,0}

\def\gsim{\, \rlap{$>$}{\lower 1.1ex\hbox{$\sim$}}\,}
\def\lsim{\, \rlap{$<$}{\lower 1.1ex\hbox{$\sim$}}\,}
\newcommand{\be}{\begin{equation}}
\newcommand{\ee}{\end{equation}}
\newcommand{\bea}{\begin{eqnarray}}
\newcommand{\eea}{\end{eqnarray}}

\begin{document}


\begin{titlepage}
\bigskip
\bigskip\bigskip\bigskip
\centerline{\Large \bf The Black Hole information problem: past, present, and future}

\bigskip\bigskip\bigskip
\bigskip\bigskip\bigskip

 \centerline
 {\bf Donald Marolf\footnote{\tt marolf@physics.ucsb.edu }}
 \medskip

\centerline{\em Department of Physics}
\centerline{\em University of California}
\centerline{\em Santa Barbara, CA 93106--9530 USA}

\bigskip\bigskip\bigskip

\bigskip\bigskip\bigskip

\begin{abstract}
We give a brief overview of the black hole information problem emphasizing fundamental issues and recent proposals for its resolution.  The focus is on broad perspective and providing a guide to current literature rather than presenting full details.  We concentrate on resolutions restoring naive unitarity.
\end{abstract}
\end{titlepage}

\baselineskip = 16pt

\section{Background}

It has been understood since the early 1970's that black holes behave in many ways like standard thermodynamic systems.  In particular, a natural black hole analog of equilibrium state is given by restricting to black holes with a so-called horizon-generating Killing field, meaning that the black hole spacetime has a symmetry whose action on the black hole horizon shifts each point in a null direction.  The first law of black hole mechanics states that exchanges of energy $dE$, charge $dQ$, and angular momentum $dJ$ with such black holes satisfy

\begin{equation}
\label{FirstLaw}
dE = T dS_{BH} + \Omega dJ + \Phi dQ,
\end{equation}
where $T, \Omega, \Phi$ may be interpreted as the temperature, angular velocity, and electric potential of the black hole.   The second law states that $S_{BH}$ does not decrease with time.  As explained by Bekenstein \cite{Bekenstein:1973ur} and Hawking \cite{Hawking:1974sw}, the quantity $S_{BH}$ should thus be interpreted as an entropy.   In the Einstein-Hilbert theory of gravity one finds $S_{BH} = A/4\hbar G$ in terms of the surface area $A$ of the black hole horizon, though with appropriate $S_{BH}$ the first law \cite{Wald:1993nt} and (at least in linearized form about a Killing horizon \cite{Wall:2015raa}) the second law are now understood to hold  in very general theories of gravity.

Up to the normalization of entropy and temperature, these results follow directly from the classical gravitational field equations.  But Hawking's 1974 quantum field theory analysis \cite{Hawking:1974sw} showed that black holes do in fact radiate thermally at a temperature proportional to $\hbar$ and set by the above normalization for $S_{BH}$. This gives a very physical verification of the thermodynamic nature of black holes, and a striking contrast from the strictly vanishing temperature of a classical black hole.  However, it also leads to the well-publicized black hole information problem.

Our purpose below is to give a brief overview of this problem, emphasizing both fundamental issues (section \ref{HR}) and recent proposals for its resolution (section \ref{proposals}).  We will not attempt to be thorough or complete, but merely to convey some general ideas.  Familiarity with the Schwarzschild geometry and basic quantum field theory is assumed throughout.   A much more thorough recent review with an extended introduction to relevant background material can be found in \cite{Harlow:2014yka}, and an intermediate treatment in \cite{Polchinski:2016hrw}.

\section{Hawking Radiation and the information problem}
\label{HR}

Despite the above similarities to standard thermal effects, conventional physics implies the Hawking effect to differ fundamentally from familiar thermal emission from hot objects like stars or burning wood.  To explain this difference, let us first observe that while free null worldlines can skim along a black hole horizon, such trajectories are inherently unstable.  Displacing the worldline toward the outside results in the particle flying away from the black hole, while a small displacement toward the inside sends the particle into the singularity.  As a result, a thin tube of null particles near the horizon experiences an exponential stretching in the radial direction as it propagates toward the future. Since all excitations act like null particles at high energies, the same is true of the near-horizon (short-distance, and thus high energy) part of the state of any quantum field.    Black holes thus act like cosmic magnifiers, stretching microscopic features of quantum states to macroscopic scales.

This stretching effect leads to Hawking radiation in a very direct way.  First recall that at sufficiently short distances any state of our quantum field will be well approximated by the vacuum.  We must then also recall that the vacuum of quantum field theory contains divergent ultraviolet correlations between spacelike separated points.  As a result, stretching the vacuum creates sizeable correlations across macroscopic distances between points outside the black hole and those within.  The stretching effect also redshifts these correlations to low energy; indeed, to the scale set by the above Hawking temperature $T$, which for 4d Schwarzschild black holes of radius $R$ turns out to be $T=\hbar c/4\pi R$.  We thus learn that -- even when they have travelled far from the black hole -- the low energy modes (at scale $T$) of the outgoing quantum field outside the black hole define a mixed quantum state.  Indeed, since each outgoing mode at scale $T$ is strongly correlated with the black hole interior but essentially uncorrelated with other such outgoing modes, the associated density matrix $\rho_{outgoing}$ must have von Neumann entropy $S(\rho_{outgoing}) = -{\rm Tr} \left( \rho_{outgoing} \ln \rho_{outgoing} \right)$ of order one bit per mode at the scale $T$. This mixed state cannot be the outgoing vacuum (which is pure), and with only a small amount of work (see e.g. \cite{Jacobson:2003vx}) can be shown to be precisely thermal in the approximation where gravitational back-reaction is neglected.    In particular, it carries energy to infinity and by energy conservation will cause the black hole to decrease in mass, and thus in radius.  The resulting shrinkage is known as black hole evaporation.  The process begins slowly since $T \propto \hbar$ but -- at least in the Schwarzschild case -- becomes more rapid as $R$ decreases and $T$ increases correspondingly.

Now, familiar thermodynamic systems like stars are also characterized by the emission of nearly thermal radiation at a time-dependent temperature $T(t)$.  But for Hawking radiation the above argument turns out to imply that any correlations between outgoing modes must be small, no matter how long the Hawking process is allowed to continue.  In contrast, in more familiar contexts, starting with a hot object in a pure quantum state and letting it radiate all of its energy forces the remains to be in the ground state.  Typical ground states are non-degenerate, so at late times the full quantum state $|\Psi\rangle$ of the system must be of the form $|\Psi\rangle = |0\rangle_{object} |\tilde \Psi\rangle_{radiation}$; i.e., the state is a product state, so that the purity of the original state implies that the radiation is also described by a pure quantum state $|\tilde \Psi\rangle_{radiation}$ and thus has vanishing entropy.  While individual photons emitted during early stages of the radiative process may be largely uncorrelated, strong correlations are thus inevitable over the entire period of radiation to the ground state.

Let us therefore return to the argument for a highly mixed outgoing Hawking state in order to explain the above-mentioned lack of correlations and to exhibit the assumptions involved.  The discussion applies to any quantum field theory propagating on a classical spacetime background.  Quantum corrections to the classical metric can also be incorporated by expanding gravity perturbatively around a classical black hole background and treating the result as an effective field theory with a cutoff.

It is useful to phrase the argument for small outgoing Hawking correlations in terms of the monogamy of quantum entanglement: If system B is strongly entangled with some system A, then its entanglement with all other qubits must be small.  This is a consequence of the famous strong sub-additivity property of quantum von Neumann entropies $S_X$ which states that for independent systems $X,Y,Z$ and their unions $XY, YZ, XYZ$ we have

\begin{equation}
\label{SSA}
S_{XYZ} + S_Y \le S_{XY} + S_{YZ},
\end{equation}
and so in particular
\begin{equation}
\label{SimpSSA}
S_Y \le S_{XY} + S_{YZ}.
\end{equation}
If $XY$ is very pure (and thus $S_{XY} \ll S_Y$), we must have $S_{YZ} \gtrsim S_Y$  for any system $Z$; i.e., no system $Z$ can significantly purify $Y$.

To apply this argument to the Hawking effect, we consider the limit of small back-reaction in which the spacetime metric evolves parametrically slowly compared with the timescale set by the black hole temperature $T$; e.g., the limit of a large Schwarzschild black hole.  As a result, we may approximate the metric as being time-independent even over appropriate times $\Delta t \gg 1/T$.  And as described above, at late such times the outgoing part of any state will be thermal and thus highly mixed.  Let us denote by system $B$ the outgoing part of the state with frequency $\sim 1/T$ that emerges from the near-horizon region over some time interval $I_1$ of width $w_1$ near some time $t_1$ such that $1/T \ll w_1 \le \Delta t$, while we call the corresponding part of the state inside the horizon system $A$.  Thus $AB$ is just the frequency $\sim 1/T$ part of the vacuum near the horizon in the interval $I_1$ and can be regarded as essentially pure ($S_{AB} \ll S_B$).  It then follows from the Araki-Lieb inequality $S_{AB} \ge |S_A - S_B|$ that $A$ itself is also thermal, with $S_A = S_B$ up to small corrections.

We then choose $C$ and $D$ to be the corresponding outgoing and ingoing parts of the state with frequency $\sim 1/T$ emerging from the near-horizon region over some other later interval $I_2$ of width $w_2$ near another time $t_2$ such that $1/T \ll w_2 \le \Delta t$ with, say, $|t_2 - t_1| \gg \Delta t$. At least according to the rules of standard quantum field theory, the systems $A,B,C,D$ must be independent since the degrees of freedom they describe can be localized in spacelike-separated regions of the spacetime at late time.  In particular, both $A$ and $D$ fall far behind the horizon and -- despite the shrinkage of the horizon as the black hole evaporates -- become well-separated from $B$ and $C$.

We now apply \eqref{SSA} to the systems $X=D$, $Y=C$, $Z=AB$.  Noting that $S_{XYZ} = S_{ABCD}$ and $S_{XY} = S_{CD}$ are small, this gives $S_{YZ} = S_{ABC} \ge S_C$.  Taking this as input and again recalling that $S_{AB}$ is small, we apply \eqref{SSA} again to the systems  $X=A$, $Y=B$, $Z=C$ to conclude $S_{BC} \gsim S_B + S_C.$  And since subaddivity requires $S_{BC} \le S_B + S_C$, we in fact have $S_{BC} \approx S_B + S_C$. Significant correlations between outgoing degrees of freedom at different times are thus forbidden and the entropy of the outgoing radiation is roughly one bit per outgoing mode at scale $T$ as claimed above.  A careful quantitative version of this argument can be found in \cite{Mathur:2009hf}, which also holds in the presence of small corrections not necessarily described by quantum field theory; see also \cite{Sorkin:1997ja,Giddings:2006sj,Braunstein:2009my}  and extensions of \cite{Mathur:2009hf} in \cite{Giddings:2012dh}.

\subsection{A straw man information problem}
\label{straw}

Extrapolating the above semi-classical reasoning to the point where the black hole evaporates completely leads to the ideas of  {\it non-unitary evolution} and {\it information loss} \cite{Hawking:1976ra}.  The point is simply that, if that black hole itself has completely disappeared then -- at least from the perspective of the outside universe -- one is left only with the above radiation.  So even if the universe began in a pure state, it would have evolved into one that is highly mixed.  This is impossible under unitary time evolution on a closed system.  And because the end state of the radiation would be largely independent of the initial state, knowledge of this final state would not suffice to deduce the initial state and information would have been lost.

Such behavior is starkly different from that of familiar quantum systems. But this observation is not yet a serious problem as there are at least two natural ways to attempt to restore unitarity.  The first is to note that during the final stages of its evaporation the black hole would be very small and curvatures at the horizon would be Planck scale.  Thus quantum gravity corrections to the above discussion would naturally become very large.  There is then no reason to expect semi-classical reasoning to be even approximately valid at that stage of the evolution.  In particular, the final state can be pure due to the initial (highly mixed) radiation being strongly entangled with the whatever results from this late-time evolution.  Such ideas have been traditionally called remnant scenarios \cite{Hawking:1982dj} with the idea that Hawking radiation might largely turn off at this stage and leave a more-or-less stable Planck-sized object with high entropy that would take a long time to decay.  However, as noted in \cite{Almheiri:2013wka}, they also include the so-called bounce scenarios of \cite{Ashtekar:2008jd,Ashtekar:2010hx,Ashtekar:2010qz}.  Classic discussions of remnants and their consequences can be found in \cite{Giddings:1993km,Giddings:1994qt}.

The second natural way to restore unitarity is to suppose that there are additional degrees of freedom not part of what was called `the external universe' above; i.e., that the system being studied was not in fact closed.  The idea here is that the ingoing parts of the quantum field (systems $A$ and $D$ above) in some sense continue to both exist and to remain separate from the external universe even after the black hole decays completely.  One may think of this as being due to the creation of a new so-called baby universe (though perhaps one that is intrinsically Planck-scale and very quantum) that separates from the external one due to the collapse to a singularity of the black hole interior (see e.g. \cite{Markov:1984ii}).  The idea is then that the evolution on the entire system (external universe + baby universe) can remain unitary.

\subsection{The information problem and the Page time}
\label{Page}

However, as argued by Page \cite{Page:1993df}, a related problem involving the Bekenstein-Hawking entropy
$S_{BH} = A/4\hbar G$ in fact arises much earlier in the evaporation process.  Since $S_{BH}$ satisfies the first and second laws of thermodynamics, it is natural to assume that number of internal states available to a black hole with given surface area $A$ is $e^{S_{BH}}$.  In particular, this number should include the states of any baby universes formed from our black hole thus far, as there is no separate term in either the first or second law to account for such offspring.  Thus $S_{BH}$ would be an upper bound on the von Neumann entropy $\rm{Tr} (\rho {\rm log} \rho)$ for any black hole.  But interchanging ingoing and outgoing modes above shows that -- so long as the black hole remains large enough for semi-classical reasoning to apply over timescales of order $1/T$ -- the von Neumann entropy of the black hole must increase by roughly one bit for each emitted Hawking photon.

This is now a real problem.  Evaporation causes the black hole to shrink, and thus to reduce its surface area.  So $S_{BH}$ decreases at a steady rate.  On the other hand, the actual von Neumann entropy of the black hole must increase at a steady rate.  But the first must be larger than the second.   So some contradiction is reached at a finite time.  For a black hole that begins in a pure state, this is readily seen to occur when $S_{BH}$ has been reduced to half of its initial value.  This is known as the Page time $t_{Page}$.  For $4d$ asymptotically-flat Schwarzschild black holes of initial radius $R_0$ one finds $t_{Page} \sim R_0^3/\ell_{Planck}^2$ in terms of the Planck length $\ell_{Planck}.$   But in any case one sees that for large $R_0$ the black hole remains far from the Planck scale at $t_{Page}$ as the area of the black hole has decreased only by a factor of $2$.  So as opposed to the case in the late-time regime discussed above, strong quantum corrections before $t_{Page}$ would be a great surprise.

\section{Resolutions?}
\label{proposals}

We now comment briefly on various proposals to resolve the above information problem.  While constraints on space prohibit an in-depth discussion of each proposal, or even a complete list of proposals, further details can be found in the references below.  This section ends with a list of remarks in section \ref{remarks} which may be useful in understanding how certain other parts of the literature relate to the ideas discussed in sections \ref{InfoLoss}-\ref{MQM}.

\subsection{No problem in principle?}
  \label{InfoLoss}
  Page's argument relies on interpreting $S_{BH}$ as a standard density of states for the black hole.  At least within string theory, strong evidence that this is the case comes both from Strominger-Vafa style counting of D-brane bound states \cite{Strominger:1996sh} and -- even better -- from gauge/gravity duality where (logarithms of) lattice gauge theory partition functions precisely match free energies $F =  E-TS_{BH}$ of bulk black holes \cite{Hanada:2008ez,Hanada:2012eg,Hanada:2014hpa}.

But it remains interesting to ask if there are other options outside string theory.  This essentially means returning to the remnant or baby universe scenarios of section \ref{straw}. While I know of no sharp argument that absolutely forbids such scenarios, they do raise issues which I briefly mention below, and on which little progress has been made over the past 30 years.  The issues are largely field theoretic in nature, and thus are not apparent in the toy model of baby universe scenarios discussed in \cite{Unruh:2012vd}.

An important point is that any remnant or baby universe resolution must be quite extreme.  Suppose that a Schwarzschild black hole of macroscopic area $A$ has some finite number of internal states $e^{S_{A}}$.  And without strong violations of either quantum mechanics or effective field theory, the arguments of section \eqref{HR} show that when a black hole evaporates from area $A_1$ to some smaller area $A_2$ its von Neumann entropy must increase by $\sim (A_1 - A_2)/4G$.  So no matter how large we choose $S_{A_2}$, for $A_1 \gg A_2 + 4GS_{A_2}$ we reach the same contradiction as in section \ref{Page} when the black hole evaporates down to area $A_2$.  As a result, we must have a truly infinite density of states at fixed black hole area, or equivalently at fixed black hole mass.  Note that the density of states must remain infinite even in the presence of an infra-red (IR) regulator.  For example, resolving the problem for black holes in asymptotically anti-de Sitter (AdS) spaces with reflecting boundary conditions requires the system to have a strictly infinite number of states at each energy $E$.

Various questions then follow.  A striking example \cite{Giddings:1993km} arises when one assumes that the divergent density of states arises only when a black hole or some other manifest remnant is present in the quantum state.  This in particular occurs if one assumes Hawking radiation to simply turn off once the black hole reaches some Planck-scale size.  Since any effective field theory treatment implies that objects of finite energy with an infinite density of states are copiously produced in any interaction with sufficient energy, one would expect any object larger than a speck of dust (and thus with rest mass energy $mc^2$ larger than the Planck energy) to rapidly transform itself into such a high-entropy remnant by some quantum tunneling process.

Even dropping the assumption of an effective field theory treatment, one notes that many features of familiar systems are dictated mainly by state counting.  But such counting arguments are ill-defined with an infinite density of states. For example, what rules would determine the probability of finding a black hole (or any other object) in a thermal ensemble?  To be concrete, we might consider the outgoing Hawking radiation produced by a large black hole.  As described above, this is largely thermal, and has the advantage that the universal coupling of gravity to all forms of energy can be expected to remove any dependence of the probability on the object's internal state.  A closely related question (see my comments in \cite{Jacobson:2005kr}) asks why $S_{BH}$ would appear in \eqref{FirstLaw} if it does not give the black hole density of states.

A classic studies of at least one form of this idea were performed by Giddings and Strominger \cite{Giddings:1988cx} and by Coleman \cite{Coleman:1988cy}, who focused on the idea of baby universes and used ideas related to effective field theory.  Despite first appearances, they showed  within this framework that such models do not in fact lead to loss of unitarity in the external universe.  Instead, the internal state of the baby universes turns out to define superselection sectors for the algebra of operators in the external universe.  Within each sector,  the external universe evolves unitarily with some fixed Hamiltonian, though with coupling constants that depend on the choice of sector.  It would be interesting to return to this analysis and clarify precisely what assumptions lead to this conclusion and whether they might be avoided in reasonable models.

Finally, we summarize the (perhaps related) boundary unitarity argument of \cite{Marolf:2008mf} which also constrains baby universe scenarios.  In essence, the argument there is that (modulo terms that vanish when the equations of motion hold) in classical gravitational theories the Hamiltonian is given by a flux integral over the spacetime boundary.  Indeed, this is a cherished feature of gravitational physics that is closely related to invariance under diffeomorphisms.   A prime example is the familiar Arnowitt-Deser-Misner Hamiltonian (see e.g. \cite{Wald:1984rg}) in asymptotically flat spacetimes, though for simplicity we assume here that the boundary is timelike (as in asymptotically AdS spacetimes).   We similarly assume that the boundary has a time translation symmetry so that the Hamiltonian is conserved; i.e., $H(t) = H(t')$ for any $t,t'$, though the time-dependent case can also be handled with a bit more work.  So if the full theory of quantum gravity is unitary in the above sense, and if the Hamiltonian remains a boundary term, then the on-shell time evolution operator $e^{iHt}$ can be constructed using only quantities on any spacetime boundary.  And since Heisenberg evolution gives
$O(t_0) =  e^{iH(t_0-t)}O(t) e^{-iH(t_0-t)}$, any information that can be obtained from an operator $O(t_0)$ on the boundary at some time $t_0$ can also clearly be obtained from $e^{iH(t_0-t)}O(t) e^{-iH(t_0-t)}$ at time $t$.  But the latter is an operator living only on the spacetime boundary at the (perhaps much later) time $t$.  It therefore seems unnatural to me to say that any information has been ``lost to another universe.''   Reconciling this argument with the baby universe scenario requires either a theory of quantum gravity where the Hamiltonian fails to be given by gravitational flux through the boundary, or a theory where the state of baby universes influences this flux.  The latter is manifestly non-local, and so should be compared with the resolutions in section \ref{EFTF} below in which effective field theory also fails but naive unitarity is restored.

\subsection{No problem in practice?}  Although taking $e^{S_{BH}}$ to be the number of black hole internal states conflicts with the simultaneous use of standard quantum kinematics and low energy effective field theory dynamics, it was argued in \cite{Susskind:1993if} that such conflicts might be acceptable if they were not visible to any single observer.  In particular, there would be no tension with any known experimental results.

This idea was called Black Hole Complementarity in analogy with Heisenberg Complementarity in quantum mechanics.  In particular, it was conjectured that experiments performed by observers who remain outside the black hole at all times would be sensitive to the finite density of states while those performed by observers who fall in would not, and that the results of each set would be separately consistent with known physics.

However, as shown in \cite{Almheiri:2012rt}, with enough sophistication one can in fact find experiments that reveal the above contradiction to a single observer.  While  one may attempt to find stronger versions of complementarity  \cite{Bousso:2012as,Harlow:2013tf} that avoid the problem raised in \cite{Almheiri:2012rt}, other single-observer issues remain \cite{Almheiri:2013hfa,Oppenheim:2014gfa}, especially when one considers suitably generic black holes \cite{BoussoTalk,Marolf:2013dba,Bousso:2013wia}.

\subsection{Effective Field Theory Fails?}
\label{EFTF}

We thus conclude that taking $e^{S_{BH}}$ to give the black hole density of states requires introducing novel physics to describe the experiences of single low-energy observers in weakly curved backgrounds.  This is a surprise, as one might have thought physics in this regime to be well understood.  Furthermore, due to the magnitude of the entropy problem described in section \ref{HR}, one might expect this new physics to have dramatic effects.  Indeed, the authors of \cite{Almheiri:2012rt} expressed a belief that any viable option would require so much drama that one may as well simply suppose that, after the Page time, quantum fields near black hole horizons are not in fact well-described by vacuum.  At the very least, they would instead be highly excited due to new unknown physics.  The set of excitations was called a ``black hole firewall,'' and might even be sufficiently strong that spacetime failed to exist in any recognizable sense in the interior of such black holes\footnote{Our discussion concerns black holes that evolve adiabatically slowly.  As a result, to good approximation a region of spacetime at any time $t$ may be modeled by a corresponding region of a stationary black hole spacetime.  The near-horizon region described above is the region corresponding to spacetime close to the Killing horizon of this comparison stationary black hole; i.e., near what we may call the ``adiabatic horizon'' (which becomes sharply defined only in the adiabatic limit).  In particular, there is no reason to associate the firewall directly with the event horizon (whose location at one time can in contrast be determined only by knowing what will fall into the black hole at all later times). }.

Since standard effective field theory leads to vacuum near the horizon as described in section \ref{HR}, the creation of a firewall requires strong novel effects; i.e., it would imply a spectacular failure of effective field theory.  Nevertheless, the computations \cite{Banerjee:2011jp,Sen:2011ba,Sen:2012cj,Bhattacharyya:2012ye} of logarithmic corrections to the density of states for extremal black holes may be taken as evidence that firewalls are indeed present around generic black holes.  The point here is that the string-theoretic density of states was shown to precisely matches the sum of the Bekenstein-Hawking entropy $S_{BH}$ of the black hole and the entropy of the surrounding radiation (which, even at the vanishing temperature of an extremal black hole, turns out to be non-zero due to the infinitely deep throat near the horizon in which the radiation lives).  So interpreting $S_{BH}$ as the entropy of the black hole itself, this agreement implies that generic states have radiation outside the black hole unentangled with the black hole interior; i.e., that there is a firewall at the edge of the black hole.

However, with an eye toward possible subtleties in this argument, working outside of string theory, or in an effort to ameliorate the problem for less generic black hole states -- say, those formed by rapid collapse of `normal' matter, whose von Neumann entropy is thus much smaller than $S_{BH}$ --
 it is clearly of also interest to ask both whether modifications of effective field theory resolve the information problem without such dramatic effects.  And to further the firewall hypothesis itself it is important to ask whether physical motivations for such dramatic effects can be found.  We briefly comment on both below.

\subsubsection{Non-violent Nonlocalty}
\label{NVNL}

Giddings has long championed the idea that the information problem requires corrections to standard low-energy effective field theory but that one should endeavor to make these as mild as possible \cite{Giddings:2011ks,Giddings:2012gc,Giddings:2013kcj}.  In particular, the corrections should be non-local in order to couple Hawking radiation to degrees of freedom that fell into the black hole long ago.  In terms of the discussion from section \ref{HR}, the idea is that systems $A$ and $C$ would then fail to be independent.

The particular approach advocated in \cite{Giddings:2011ks,Giddings:2012gc,Giddings:2013kcj} is to maintain both smoothness of the state and approximate validity of standard effective field theory near the horizon, but for the new corrections to allow the black hole interior to interact with degrees of freedom outside the horizon.  Confining these corrections to a parametrically small region near the horizon would require infalling observers to encounter excitations of parametrically large energy, so the corrections are instead assumed to extend a sizeable distance outside -- say, at least for Schwarzschild black holes,  a proper distance of order the black hole radius $R$.  One would then expect such corrections to transfer energies of only order $1/R$, comparable to the Hawking temperature\footnote{Interestingly, the models of such corrections in \cite{Giddings:2014nla} in fact behave like order-one modifications of the metric.  They thus deflect all null particles through the same angle, and so in fact transfer large momenta to particles of high momentum. One expects that this can be avoided in more sophisticated models, and \cite{Giddings:2017mym} suggests that this will be so in the models studied there.}. But there could still be significant effects on radiation with frequency $\sim 1/R$, and in particular on the gravitational wave signals produced by merging black holes.

Such non-violent nonlocality (NVNL) scenarios come in two basic flavors.  In the first, the extra NVNL interaction transfers entanglements to the outgoing Hawking photons once they have reached a distance of order $R$ from the horizon as described above.  As a result, there is an opportunity for some apparatus, introduced by an intrepid experimenter, to interact with the Hawking modes as they travel from the near-horizon region to the point of entanglement transfer.  Complications then arise \cite{Almheiri:2012rt} if the apparatus absorbs the Hawking photons (so that the extra interaction at the scale $R$ can act only on the outgoing vacuum), or even \cite{Almheiri:2013hfa} if the apparatus simply transforms the state of the Hawking modes into one other than that expected from the argument in section \ref{HR}.   Indeed, \cite{Almheiri:2013hfa} concludes (see its footnote 14) that contradictions with unitarity can be avoided only by supposing that any instructions sent to the apparatus from far away somehow affect the extra NVNL interaction in a compensating manner.

In the second version of NVNL, the original photons are left untouched but the corrections produce an additional flux of energy which can carry entanglements away from the black hole.  The idea is then for this new flux to decrease the black hole's von Neumann entropy faster than the original Hawking flux causes it to increase\footnote{In 1+1 dimensions, the well-known infra-red (IR) divergences for massless fields imply that the vacuum state of such fields contains strong entanglements between well-separated low-energy degrees of freedom.  These can be used to transfer large quantities of entanglement with negligible energy \cite{Wilczek1992,Hotta:2015yla}. In particular, as described in \cite{Almheiri:2013wka} this effect appears to be involved in the system analyzed in \cite{Ashtekar:2008jd,Ashtekar:2010hx,Ashtekar:2010qz}. But there are no known examples of similar effects in higher dimensions.  }.

However, thermodynamics turns out to place fundamental constraints \cite{Almheiri:2013hfa} on this second class of scenarios.  So it is no surprise that ideas along these lines have been found to require either a black hole density of states smaller than $S_{BH}$ by a factor of order $1$ \cite{Giddings:2013vda}, or a change in the so-called grey body factors to make the system much more of an ideal black body \cite{Giddings:2015uzr}.  We note that the latter scenario leads to correspondingly large changes in the absorption cross section for radiation of frequency $1/R$, and thus perhaps in the gravitational wave signals mentioned above.  The former scenario, and in particular the models in \cite{Giddings:2017mym}, also seem likely to affect such gravitational wave signals though the relevant calculations have yet to be performed.

Unfortunately, even such effects do not suffice to resolve the information problem for generic black hole states in equilibrium with their environment \cite{Almheiri:2013hfa}.  This is essentially due to further fundamental issues in thermodynamics:  When two systems exchange energy, the standard relations between entropy and entanglement are consistent only when all modes of energy transfer can in principle transfer entanglement as well. Indeed, in both versions of the NVNL proposal the fundamental issue appears to be that the transfer of energy from the black hole (via the normal production of Hawking radiation) is a fundamentally separate process from the desired transfer of information.

\subsubsection{Equating the interior and exterior}

Rather than relate the interior and exterior via a dynamical interaction, one might also simply posit new kinematics in which the systems called $A,D$ in section \ref{HR} are somehow controlled by the same degrees of freedom as those for the outgoing Hawking radiation \cite{Bousso:2012as,Nomura:2012sw,Papadodimas:2012aq,Nomura:2012cx,Harlow:2013tf,Susskind:2013tg}  (though we note that \cite{Bousso:2012as} argued against this idea). In general, this redundancy is visible to infalling observers who see this as a large departure from known physics.  In particular, as discussed in \cite{Almheiri:2013hfa}, they see phenomena that appear to constitute quantum cloning (and which are thus forbidden in standard quantum mechanics \cite{Wootters:1982zz,Dieks:1982dj}).  In addition, interacting with radiation already emitted at some time $t$ will generally create a firewall even if one was not present originally \cite{Almheiri:2013hfa,Bousso:2013wia,Bousso:2013ifa}.  There are, however, special situations \cite{Maldacena:2013xja} where these pathologies do not arise.   The so-called ER=EPR proposal expresses the hope that these special examples may indicate more general resolutions of the black hole information problem.   But ER=EPR alone cannot resolve the problem for generic black holes \cite{BoussoTalk,Marolf:2013dba,Bousso:2013wia,Bousso:2013ifa}, and thus also does not by itself suffice for black holes that have been evaporating for a Page time.  Proponents of ER=EPR thus generally suggest \cite{Susskind:2014moa} that a full resolution also requires quantum mechanics to be replaced by some non-linear theory (see e.g. section \ref{MQM} below).

\subsubsection{Sources of Drama}

If dramatic effects near the horizon are required to resolve the information problem, then it is crucial to understand whence these might arise.  Some ideas \cite{Giveon:2012kp,Silverstein:2014yza} are under investigation within string theory, though they have not yet been clearly shown to suffice.    There are also arguments in loop gravity that quantum gravity does in fact lead to larger departures from effective field theory than one might expect.  In particular, \cite{Rovelli:2014cta,Haggard:2014rza,Christodoulou:2016vny} suggest that matter collapsing into a 4d Schwarzschild black holes could undergo a quantum transitions to a thin shell expanding from a white hole on timescales of order $t_{bounce} \sim R_0^2/\ell_{Planck}$ that are much shorter than the Page time $t_{Page}$.

However, this latter proposal cannot actually resolve the information problem in the form discussed here.  If the above black hole has $e^{S_{BH}}$ internal states with $S_{BH} \sim R_0^2/\ell_{Planck}^2$, then we may suppose that it begins in a maximally-entangled state with von Neumann entropy $\sim A/4G \sim R_0^2/\ell_{Planck}^2$.  For simplicity, we suppose that the ancillary system (``the ancilla") with which it is entangled is far away so we may ignore subtleties associated with short-distance QFT vacuum entanglement in the region near the black hole.  Unitarity then forbids evolution to a system with smaller entropy without interaction with the ancilla.  But if the black hole evaporates in a time $t_{bounce}$ by emitting normal radiation into the region near a horizon of area $\sim R_0^2$, the maximum entropy that can result is given by thermal radiation of total energy $E \sim R_0/\ell_{Planck}^2$ filling a region of volume $V \sim A t_{bounce} \sim R_0^2/\ell_{Planck}$.  And in 4 spacetime dimensions this thermal radiation has entropy $S \sim E^{3/4} V^{1/4} \sim R_0^{7/4} \ell_{Planck}^{-7/4}$ which is much less than $S_{BH} \sim R_0^2/\ell_{Planck}^2$ for $R_0 \gg \ell_{Planck}$. So if $S_{BH}$ describes the number of internal black hole states the above transition will be forbidden.

This argument should remind the reader of the thermodynamic constraints on NVNL from section \ref{NVNL}.  Indeed, in many ways the bounce scenario acts like a more violent form of NVNL with extra energy flux.  It is thus subject to the same constraints.  In particular, as with some versions of NVNL, proponents of the above bounce scenario might suppose that the actual number of black hole states is smaller than $S_{BH}$.  But this leads to issues regarding equilibrium and the first law similar to those described in section \ref{InfoLoss}. Furthermore, having all of the energy come out at the bounce time $t_{bounce}$ gives a highly unstable scenario.  The point here is that  gravitational interactions of the outgoing shell from the white hole with even a small amount of infalling matter (e.g., a photon from the cosmic microwave background) can cause the shell to recollapse into a new black hole \cite{PhysRevLett.33.442,Barcelo:2015uff}.  For stability, the release of energy should be spread as evenly as possible between $t_{bounce}$ and the initial formation of the black hole, requiring a constant stream of radiation to be emitted from near the black hole horizon. In particular, it differs from Hawking radiation in that it is {\it not} experienced as vacuum by incoming observers.  Indeed, if the emitting region is exponentially close to the horizon as in \cite{Christodoulou:2016vny},  infallers will then experience the radiation just now being emitted as being blueshifted to exponentially high energies.  So in the end such scenarios again greatly modify effective field theory and -- if they do in fact follow from full quantum gravity calculations -- may be viewed as providing rationales for firewalls.

\subsection{Modify Quantum Mechanics?}
\label{MQM}

The reasoning of section \ref{Page} made heavy use of the kinematic framework of standard quantum mechanics.  It could thus be that the information problem points to a need to replace this structure with something more general.  There is of course a long history of attempts to modify quantum mechanics, and it is not something to be undertaken lightly. Indeed, it is the most general framework that satisfies a set of cherished physical postulates \cite{Kapustin:2013yda}.

Neverthless, two proposals for generalizations have been studied in the hopes of resolving the information problem. The first is the black hole final state proposal \cite{Horowitz:2003he}.  While this proposal allows a system forming a black hole to be in any density matrix $\rho_{initial}$, it requires the density matrix $\rho_{final}$ at the future spacetime boundary to be pure at the black hole singularity.  Probabilities for strings of events (with each event represented by a projection operators $P_{i}$) are then calculated according to the rule
\begin{equation}
probability = \rm{Tr} \left(\left[\prod P_i \right]\rho_{initial} \left[\prod P_i \right]^\dagger \rho_{final} \right).
\end{equation}
In particular, \cite{Lloyd:2013bza} showed that this proposal can avoid the specific issues raised in \cite{Almheiri:2012rt,Almheiri:2013hfa}, though it nevertheless entails significant implications for the experiences of certain infalling observers \cite{Lloyd:2013bza,Bousso:2013uka}.

The second proposal makes fundamental changes to the notion of quantum mechanical observable.  In textbook quantum mechanics, observables are Hermitian linear operators on the Hilbert space.  But \cite{Papadodimas:2012aq,Papadodimas:2013wnh,Papadodimas:2013jku,Papadodimas:2013kwa,Papadodimas:2015xma,Papadodimas:2015jra} propose\footnote{Related ideas
appeared in~\cite{Nomura:2012sw,Nomura:2013gna,Nomura:2012ex} and~\cite{Verlinde:2012cy,Verlinde:2013uja,Verlinde:2013vja,Verlinde:2013qya}.} to replace this structure with what one may call state-dependent operators, where the choice of linear operator corresponding to a given physical observable is influenced by the quantum state on which it will act.  In effect, this makes quantum mechanical observables non-linear.  While it remains to understand whether a fully consistent framework can be built in this way, it has been shown \cite{Marolf:2015dia} that implementations sufficient to resolve the information problem will predict massive violations of the standard Born rule for computing probabilities.  In particular, it requires the existence of normalized states  $|\alpha \rangle, |\beta \rangle$ for which the norm of $|\alpha \rangle - |\beta \rangle$ is small but where $|\alpha \rangle, |\beta \rangle$ represent mutually exclusive physical alternatives.  One may then discuss \cite{Marolf:2015dia,Raju:2016vsu} the sense in which such violations are observable.

\subsection{Further Remarks}
\label{remarks}

We end with a few brief remarks clarifying how the above points relate to certain parts of the black hole information literature.

{\bf Scrambling:}  Our treatment has emphasized the Page time $t_{Page}$, which for
for $4d$ asymptotically-flat Schwarzschild black holes of initial radius $R_0$ takes the form $t_{Page} \sim R_0^3/\ell_{Planck}^2$.  At this time, an initially pure-state black hole becomes maximally entangled.  The black hole itself is then described by a density matrix equivalent to that of a microcanonical ensemble. According to standard quantum mechanics, we may thus think of the black hole after this time as being in a randomly selected state.

However, it has been argued \cite{Hayden:2007cs,Sekino:2008he} that black holes take on some properties of random states at a much earlier time $t_{scrambling} \sim R_0 \ln \left(R_0/\ell_{Planck}\right)$.  It is thus an interesting question \cite{Almheiri:2012rt} whether the information problem and any of its proposed resolutions might manifest themselves on this earlier timescale after a black hole first forms.

{\bf Fuzzballs and Fuzzball Complementarity:}   The so-called fuzzball program \cite{Mathur:2005zp,Bena:2007kg,Balasubramanian:2008da,Skenderis:2008qn,Mathur:2008nj,Chowdhury:2010ct,Bena:2013dka} has been a very active approach to resolving the information problem.  The idea is that, in the presence of would-be black holes, string theory may provide novel physics even at macroscopic distance scales.  But in practice the actual ways that this program might resolve the information problem can be discussed as above.  In particular, let us consider the fuzzball complementarity idea of \cite{Mathur:2012jk,Mathur:2013gua}, which suggests that the approach of an infalling observer creates new regions of empty vacuum from a would-be firewall.

The creation of such vacuum regions entails sizeable corrections to effective field theory dynamics, but nevertheless is not expected to remove the entire firewall.  The new regions of vacuum are associated with deforming the horizon of the original black hole to the larger one that results after the infaller has entered.  In particular, they remove  from the firewall only those modes shorter than the distance scale set by the growth of the horizon. And since they arise from back-reaction, the new regions of vacuum are small unless the infaller is very massive.  The details may be modeled using calculations like those in \cite{Amsel:2007cw}.

One may thus think of fuzzball complementarity as hybrid proposal combining modifications of effective field theory with a weakened firewall.  The firewall is no longer Planck scale, so in a theory with only gravitational-strength interactions what remains might be considered a small effect.  But it still contains excitations at energies far greater than those set by spacetime curvature, and a back-of-the envelope calculation (which, unfortunately, is still too long to include here) based on \cite{Amsel:2007cw} suggests that with standard model interactions it would still destroy a human falling into e.g. the black hole at the center of the Milky way associated with Sagittarius A*.

{\bf Soft Graviton Effects:} It has been argued that long-wavelength gravitons can give rise to an infinite number of conserved charges which preserve an infinite amount of information outside black holes \cite{Strominger:2013jfa}.   It has then been proposed that this could give a new perspective on the information problem  \cite{Hawking:2016msc,Hawking:2016sgy}.   But I know of no suggestion for how precisely this would affect the entanglement issues discussed above.  Furthermore, the arguments in  \cite{Hawking:2016msc} are easily transcribed to the AdS context by using bulk surfaces with timelike segments that run along the AdS boundary for long times and interpreting the associated conserved charges as values of appropriate components of the boundary stress tensor of the AdS space.  On the other hand, in that context the AdS scale $\ell$ provides an effective infrared cut-off, reducing the effect to a finite amount of information which is known to be parametrically smaller than the Bekenstein-Hawking entropy (which is parametrically large in powers of the inverse Planck length).  So at least in the AdS context, no such effect can suffice.

\section{The future of black hole information}

The final resolution of the information problem is difficult to predict.  But much has already come from the search for insight. For example, in addition to the specific ideas described above, we remind the reader that the AdS/CFT correspondence \cite{Maldacena:1997re} (and gauge/gravity duality more generally) was discovered through related work.  And more recently the firewall hypothesis helped to motivate connecting such dualities to quantum error correction on a code subspace \cite{Almheiri:2014lwa}, a set of ideas whose implications are now just beginning to be explored.  The sharp focus into which it brings fundamental issues makes black hole information a powerful catalyst for new ideas, and suggests an exciting future for further studies of gravity, effective field theory, and quantum information.

\section*{Acknowledgements}
It is a pleasure to thank Abhay Ashtekar, Ahmed Almheiri, Steve Giddings, Jim Hartle, Daniel Harlow, Patrick Hayden, Gary Horowitz, Ted Jacobson, David Lowe, Juan Maldacena, Samir Mathur, Don Page,  Kyriakos Papadodimas, John Preskill,  Suvrat Raju, Carlo Rovelli, Eva Silverstein, Rafael Sorkin, Mark Srednicki, Douglas Stanford, Andy Strominger, James Sully, Lenny Susskind, Larus Thorlacius, Bill Unruh, Erik Verlinde, Hermann Verlinde, Bob Wald, Aron Wall and and many others for countless discussions of black hole information over many years. I especially thank Joe Polchinski for his part in our collaborations and for his constant support.  This work was funded in part by the U.S. National Science Foundation under grant number PHY15-04541 and also by the University of California.

\providecommand{\href}[2]{#2}\begingroup\raggedright

 \end{document}